\documentclass[amssymb,amsmath,superscriptaddress,preprint]{revtex4}
\usepackage[]{graphicx}
\usepackage{dcolumn}
\usepackage{bm}%
\begin{document}
\title{Role of thermal expansion heterogeneity in the cryogenic rejuvenation of metallic glasses}
\author{Baoshuang Shang}
\affiliation{Beijing Computational Science Research Center, Beijing 100193,  China}
\affiliation{Univ. Grenoble Alpes, CNRS, LIPhy, 38000 Grenoble, France}
\author{Pengfei Guan}
\email{pguan@csrc.ac.cn}
\affiliation{Beijing Computational Science Research Center, Beijing 100193, China}
\author{Jean-Louis Barrat}
\email{jean-louis.barrat@univ-grenoble-alpes.fr}
\affiliation{Univ. Grenoble Alpes, CNRS, LIPhy, 38000 Grenoble, France}
\begin{abstract}
   Cryogenic rejuvenation in metallic glasses reported in Ketov et al 's experiment (Nature(2015)524,200) has attracted much attention, both in experiments and numerical studies.  The atomic mechanism of rejuvenation has been  conjectured to be related to the heterogeneity of the  glassy state, but the quantitative evidence is still elusive. Here we use molecular dynamics simulations of a model metallic glass to investigate the heterogeneity in the local thermal expansion. We then combine the resulting spatial distribution of thermal expansion with a continuum mechanics calculation to infer the internal stresses caused by a thermal cycle. Comparing the internal stress with the local yield stress, we prove that the heterogeneity in thermo mechanical response  has the potential to trigger  local shear transformations, and therefore to induce rejuvenation during a cryogenic thermal cycling.
\end{abstract}
\maketitle
\section{Introduction}

Glasses are solid materials, disordered at small scale, but homogeneous and isotropic at large scales. It has been understood for some time, however, that their elastic properties are not uniform if they are computed on mesoscale. This heterogeneous elasticity has been characterized by a number of simulations of model glasses, more or less realistic, and also by a variety of experimental tools\cite{PhysRevLett.93.175501,PhysRevE.87.042306,wagner2011local,PhysRevLett.106.125504}. The heterogeneity in elastic properties plays a key role for understanding vibrational spectra, thermal transport, but also far from equilibrium behavior such as plastic deformation. Macroscopic, plastic deformation proceeds via microscopic rearrangements, the ``shear transformation", that take place preferentially on ``weak spots" with typically low elastic properties.  A large plastic deformation, if it can be performed homogeneously, tends to ``rejuvenate" the system, by bringing it to states of higher internal energy. 

Recently, it was discovered that an alternative to this mechanical rejuvenation could be obtained, in some metallic glasses (MGs),  by thermal cycling towards low temperatures \cite{ketov2015rejuvenation}. This is quite surprising, as from a purely thermal point of view one would expect that bringing the system to a lower temperature does not populate higher energy states, and should only result in slow aging. One possible interpretation of the effect, suggested by Hufnagel \cite{hufnagel2015metallic}, is that the rejuvenation is due to the creation of internal stresses during the temperature cycling.  The origin of such stresses could be twofold: local heterogeneities in the temperature field due to the very fast heating process which were investigated by some of us in a recent molecular dynamics study \cite{shang2018atomic}. Alternatively, an heterogeneous thermal expansion coefficient, similar to the heterogeneity in elastic constants, would also create internal stresses even if the temperature field is  homogeneous.  If the corresponding  mechanical stresses are large enough, they would generate plastic activity internal to the material, resulting in rejuvenation. Molecular dynamics simulations, performed on relatively small samples and using very fast cycling rates combined with artificial thermalisation methods, do not allow us to precisely quantify the origin of the internal stresses and rejuvenation induced by the cycling process. 

In this work we take an alternative route that bridges the gap between molecular dynamics and experiments, by using the molecular dynamics simulations to obtain quasi equilibrium properties and assessing the importance of the induced mechanical stresses by using a continuum description. 
Our first step is to quantify the heterogeneity in thermal expansion. We therefore present the first quantitative study of the local thermal expansion in a model metallic glass, putting the thermoelastic properties of glasses on the same footing as their elastic properties. We use the resulting spatial distribution of the thermal expansion as an input in a continuum model that allows us to estimate the magnitude of the stresses generated by differential thermal expansion. These stresses are finally compared to the local yield stresses that must be overcome in order to trigger plastic activity.  Our main conclusion is that, while the differential thermal expansion is indeed capable of generating plastic activity and thermal rejuvenation upon cryogenic cycling, the effect is unlikely to be observable at the scale of molecular dynamics simulations.

\section{The local thermal expansion coefficient}
The thermal expansion coefficient quantifies the  volume strain caused by a change in temperature, and can be written for a macroscopic system as 
\begin{equation}
  \alpha \equiv \frac{1}{V}\frac{dV}{dT} \big |_{p} \approx \frac{\epsilon_v}{\Delta T} \big |_{p}
  \label{eqn:1}
\end{equation}
where $\epsilon_v$ is the volume strain induced by the temperature change  $\Delta T$. 
In metallic glasses,   its  order of magnitude  is around $10^{-5}\text{ K}^{-1}$ \cite{kato2008relationship}.

In order to define a local thermal expansion coefficient, we use the coarse graining approach originally  proposed by Goldhirsch and Goldenberg \cite{PhysRevE.80.026112,goldhirsch2002microscopic}, that allows one to define continuum fields in the sense of hydrodynamics or elasticity, starting from 
atomic positions. The same approach was used previously to define local elastic constants, with results consistent with those obtained with different methods. 
  The  approach starts with the definition of a meso scale displacement field  $ \mathbf{u}(\mathbf{r},t) $ \cite{goldhirsch2002microscopic}
  from the atomic positions:
  \begin{equation}
   \mathbf{u}(\mathbf{r},t) \equiv \frac{\sum_{i}m_{i}\mathbf{u}_{i}(t)\phi[|\mathbf{r}-\mathbf{r}_{i}(t)|]}
  {\sum_{j}m_{j}\phi[|\mathbf{r}-\mathbf{r}_{j}(t)|]}
  \label{eqn:3}
\end{equation}
where $\mathbf{u}_{i}(t)$ is the displacement of atom $i$ at time $t$, starting from a reference position, and $\phi(x)$ is the coarse graining function, in this paper we choose a  Gaussian  $\phi(x) \equiv \frac{1}{v_{eff}} e^{-\frac{r^2}{2\sigma^2}}$, with $\sigma$ the coarse graining scale and $v_{eff}= (4\pi \sigma^2)^{3/2}$.

The local strain is calculated, under the assumption of small deformations, as 
$\epsilon_{ij}(\mathbf{r})=\frac{1}{2}(u(\mathbf{r})_{j,i}+u(\mathbf{r})_{i,j})$. We obtain 
 the local thermal expansion coefficient (LTEC) from  the local volumetric strain as
\begin{equation}
  \alpha(\mathbf{r}) = \frac{\epsilon_v(\mathbf{r})}{\Delta T} \big |_{p}
  \label{eqn:2}
\end{equation}
where $\alpha(\mathbf{r})$ is the LTEC at position $\mathbf{r}$, and $\epsilon_v(\mathbf{r})$ is local volumetric strain
 caused by the temperature change $\Delta T$.

We apply this procedure to a  model metallic glass $\text{Cu}_{50}\text{Zr}_{50}$ described by an embedded-atom method(EAM)  potential \cite{mendelev2009development}. The molecular dynamics(MD) simulations were conducted by an open source classical MD software: LAMMPS\cite{plimpton1995fast}.
The simulation samples contained 8000 atoms, and metallic glass samples were obtained by quenching from 2000 K to 1 K with two  different quenching rates, $10^{9}\text{ K/s}$ and $10^{13}\text{ K/s}$, respectively.
We collected configurations at 1 K , and then reheated the samples to 100 K Configurations at 1 K and 100 K were used to calculate the local thermal expansion coefficients  from the coarse-grained volumetric  strain.
During all the process, the system was controlled in pressure and temperature (NPT ensemble) using a Nose-Hoover thermostat\cite{nose1984unified} and a Parrinello-Rahman Barostat\cite{parrinello1981polymorphic}, which was always maintained at 0 bar.
\begin{figure}[!htpb]
  \begin{center}
	\includegraphics[width=1.0\textwidth]{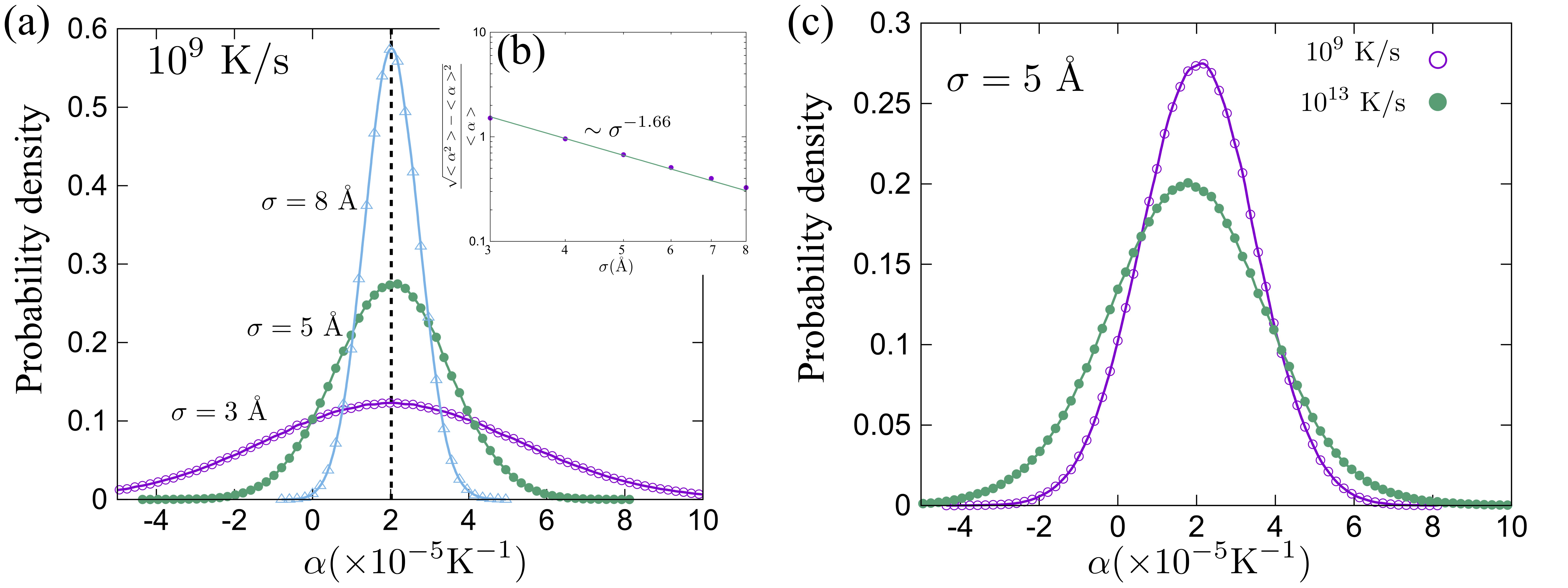}
  \end{center}
  \caption{Probability distribution of local thermal expansion coefficients($\alpha$), (a) LTEC for different coarse-graining scales in  the sample quenched at $10^{9}$ K/s, the vertical dashed line shows the macroscopic  thermal expansion coefficient obtained for the whole sample, (b) the variance of LTEC displays as a function of coarse grain size $\sigma$ which shows a power law decays with coarse grain size, (c) LTEC for samples with  different thermal histories: quenching rate $10^{9}$ K/s (hollow circle) and $10^{13}$ K/s (solid circle)}
  \label{fig:1}
\end{figure}
As shown in Fig.\ref{fig:1}(a),(b), the probability distribution function of the LTEC is typically Gaussian, with a standard deviation that increases as the coarse graining size decreases,  which is qualitatively consistent with  similar calculations of the local elastic moduli \cite{PhysRevE.80.026112,PhysRevE.87.042306}. 
 As the coarse grain size decreases, the thermal response, similar to the mechanical response, becomes heterogeneous  \cite{PhysRevE.80.026112}. 
 Interestingly,  Fig.\ref{fig:1}(c) shows that the macroscopic  thermal expansion coefficient is only weakly dependent on thermal history, but its probability distribution function  is much more sensitive to the quenching rate. High quenching rates result in higher energy states, with more heterogeneity in thermal properties.

 In the following we will be particularly interested in discussing the possibility that the stress field induced by the heterogeneous thermal expansion induces shear transformations (ST), i.e. localized irreversible plastic events that are  recognized as the elementary building blocks of plasticity in  amorphous materials\cite{tanguy2006plastic,nicolas2017deformation, hufnagel2016deformation}. Previous work on similar systems indicates that the shear transformations involve a few tens of atoms\cite{PhysRevLett.100.255901,fan2014thermally}, hence we will concentrate on results obtained for a coarse graining size $\sigma =5 \text{ \AA}$, with a coarse graining volume that contains around 30 atoms. Results obtained for different coarse graining sizes, and the influence of the value of $\sigma$,  are also investigated in the discussion section.  

\section{Internal stress by thermal expansion heterogeneity}
When the metallic glasses are heated, due to the heterogeneity of LTEC,  internal stresses are  generated by the mismatch in the local thermal deformation.
Extracting these thermal stresses   directly from molecular dynamics simulations is not easy, in particular as the system can also evolve during thermal cycling due to local, thermally activated processes that have no relation to the differential thermal expansion.
In order to estimate the role of the latter, we will adopt an alternative approach based on the assumption that the material remains a linear elastic one and can be described, at the coarse grained level,  using continuum theory, as previously demonstrated in experiments \cite{wagner2011local} and simulations \cite{PhysRevE.80.026112}.
In the following we will use the simplifying assumption that the only heterogeneity is the one of the thermal expansion, while elastic properties (bulk and shear modulus) are uniform. While this is clearly a simplification, we could not detect any correlation between heterogeneity in local elastic constants and in thermal expansion. We also note that, on the scale under consideration,   the variance  of the local elastic moduli is small compared to the mean  (see Fig.\ref{fig:app1} in Appendix), so that we expect this assumption to provide at least a correct order of magnitude of the local stresses.

\begin{figure}[!htpb]
  \begin{center}
	\includegraphics[width=0.8\textwidth]{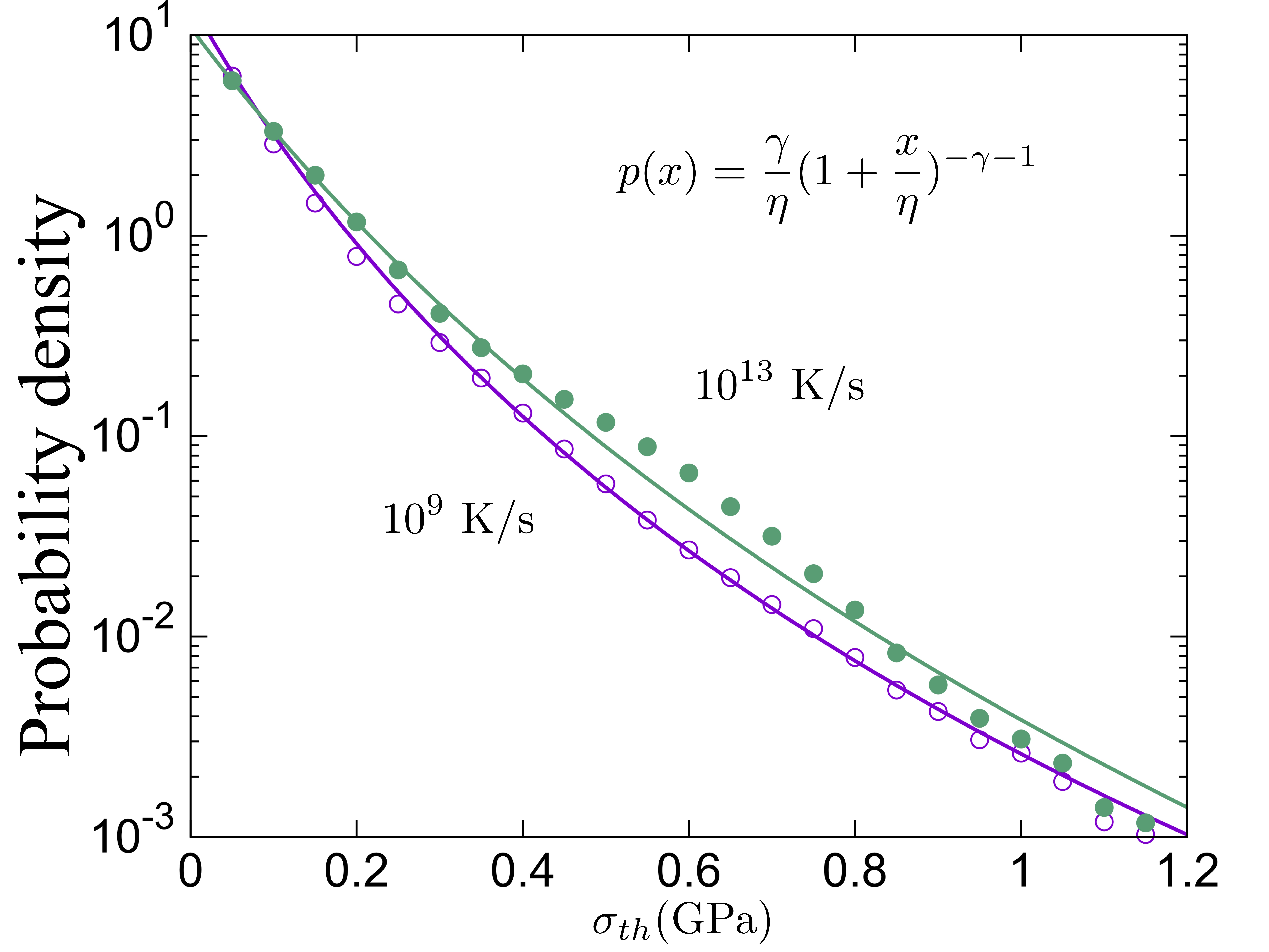}
  \end{center}
  \caption{probability distribution of thermal internal stress for different thermal history sample, the solid curve is fitted by Lomax distribution\cite{wiki:lomax}, here for numerical calculation,G=26.47,24.59 GPa;B=108.92,107.25 GPa for $10^{9}$ K/s and $10^{13}$ K/s, respectively which is also shown in Appendix and $\Delta T=100 \text{ K}$}
  \label{fig:2}
\end{figure}

With these assumptions, the effective external stress generated  by the local thermal expansion is $K \Delta T \alpha(\mathbf{r}) \delta_{ij}$ where $K$ is the bulk modulus, $\Delta T$ is the temperature  change and $\alpha(\mathbf{r})$ is the LTEC at location $\mathbf{r}$.  $\delta_{ij}$ the Kronecker symbol.
At  mechanical equilibrium, the elastic  stress would balance  the thermal stress $\sigma_{ij}$ as 
   \begin{equation}
	 \partial_j \sigma_{ij}+K\Delta T \partial_i \alpha(\mathbf{r}) =0
	 \label{eqn:4}
  \end{equation}
   using small deformation conditions and linear elasticity, the equations for the displacement field are
\begin{equation}
  G \partial_j \partial_j u_{i}+(K+\frac{1}{3}G) \partial_i\partial_j u_{j} + K\Delta T  \partial_i \alpha(\mathbf{r})=0
  \label{eqn:5}
\end{equation}
where $u_{i}$ is the $i$ component of the  displacement field at location $r$. Since the thermal expansion external force is a conservative vector field, we search  the solution for $u_i$ in the form $u_{i}=\partial_i \Phi_{,i}$  where $\Phi$ is the potential field. Equation \ref{eqn:5} can be simplified as
\begin{equation}
\partial_i\left( (K+\frac{4}{3}G)\partial_j\partial_j \Phi +K\Delta T \alpha(\mathbf{r})\right)=0
  \label{eqn:6}
\end{equation}
Integrating in a finite system with  periodic boundary conditions,  equation \ref{eqn:6} can be written as
\begin{equation}
  (K+\frac{4}{3}G)\partial_j\partial_j\Phi+K\Delta T (\alpha(\mathbf{r})-\bar \alpha)=0
  \label{eqn:7}
\end{equation}
where $\bar \alpha \equiv \frac{\int \alpha(r) dV}{V}$. Equation \ref{eqn:7} can be solved numerically using as an input the LTEC obtained from the MD simulations, and the resulting  elastic  stress can be calculated from the potential field $\Phi(r)$.  We characterize the resulting stress tensor using the  von Mises stress \cite{Fung1994First} $\sigma_{th}= \frac{\sqrt{2}}{2}\left((\sigma_1-\sigma_2)^2 + (\sigma_2-\sigma_3)^2 +(\sigma_3-\sigma_1)^2 \right)^{1/2}$ where $\sigma_i$ are the eigenvalues of the tensor. As shown in Fig. \ref{fig:2}, the probability distribution function $p(\sigma_{th} )$  shows a long tail distribution which we have fitted by the  expression $p_L(x)= (\gamma/\eta)(1+x/\eta)^{-\gamma-1}$  (Lomax distribution\cite{wiki:lomax}). We do not have any particular justification for this fit, which is only used for convenience in the following. 
\section{Probability of triggering an atomic rearrangement}
If the internal stress is larger than the local yield stress, we expect that thermal cycling will  trigger plastic activity.
In order to estimate the corresponding density of events, we first need to estimate the distribution of the local yield stresses in the material. We employ  the method
described  by Patinet and coworkers \cite{PhysRevLett.117.045501} sometimes described as  ``frozen matrix" method \cite{PhysRevE.87.042306,puosi2015probing,PhysRevLett.117.045501}.
In a low temperature system, a sphere of finite radius $r_c$ is deformed in different directions while imposing a purely affine deformation of the surrounding system, until a local plastic instability is observed, in the form of a jump in the stress-strain curve.
The von Mises stress just before the first jump in the stress strain curve is used to define a local yield stress. The  coarse graining function implicit in this approach is a discontinuous one, which separates the affinely deforming medium from the rearranging sphere.  In order to match this approach  with the one used  in computing the local thermal expansion, which uses a continuous coarse graining function, we impose that the  second moment of the two coarse graining functions are matched. This is achieved   for  $r_c\approx 2\sigma$. In the following, we use $r_c=10\text{ \AA}$, comparable to the coarse graining size $\sigma=5 \text{ \AA}$ used to obtain the thermal internal stress in the previous section.

\begin{figure}[!htpb]
  \begin{center}
	\includegraphics[width=0.8\textwidth]{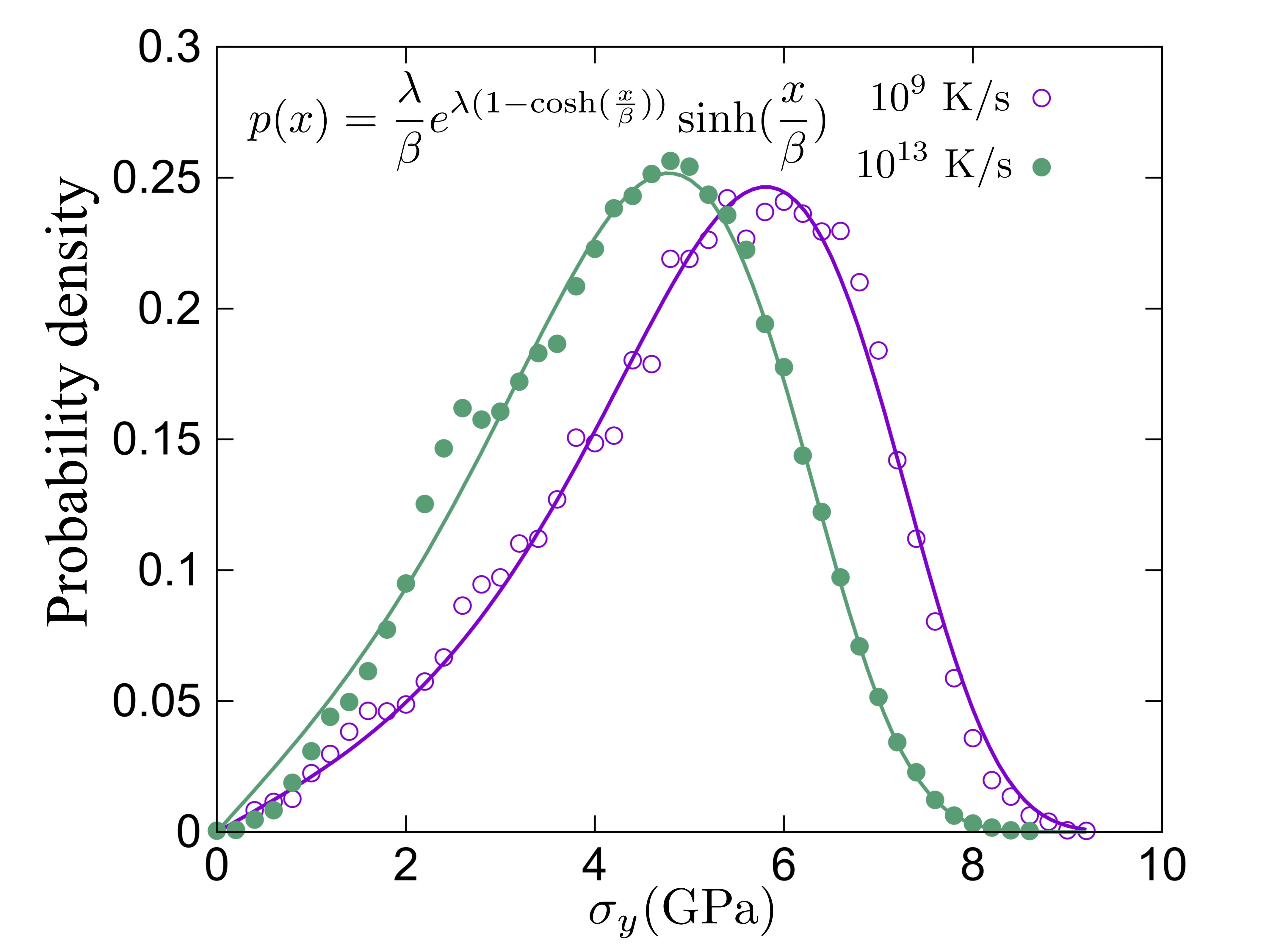}
  \end{center}
  \caption{Probability distribution of the local yield stress for different quench histories. The  solid circles correspond to the  ``fast quench"  $10^{13} \text{ K/s}$, hollow circles to the  ``slow quench" $10^{9} \text{ K/s}$. The solid curves are the  fit by  hyperbolic probability distributions. }
  \label{fig:3}
\end{figure}

The probability distribution function of the yield stress, shown in Fig.\ref{fig:3}, is well fitted by an hyperbolic probability distribution function of the form:
\begin{equation}
 p_H(x) =  \frac{\lambda}{\beta} e^{\lambda(1-\cosh(\frac{x}{\beta}))}\sinh(\frac{x}{\beta})\label{eqn:hyperbolic}
\end{equation}
Again we use this distribution for convenience, without any particular theoretical justification.
 Knowing the probability distribution of the yield stress and the one of the thermal stress, and assuming that these two quantities are statistically independent, we can estimate the 
probability that the thermal stress triggers a local plastic event within the coarse grain size, $p(\sigma_{th}>\sigma_{y})$:
\begin{equation}
   p(\sigma_{th}>\sigma_{y}) = \int_{0}^{\infty}\int_{0}^{\infty} p(\sigma_{th},\sigma_{y}) \Theta(\sigma_{th}-\sigma_{y})\mathrm{d}\sigma_{th} \mathrm{d}\sigma_{y}
  \label{eqn:8}
\end{equation}
$\Theta(x)$ is Heaviside function, $p(\sigma_{th},\sigma_{y})$ is the joint probability of the thermal stress and yield stress within coarse grain size.  We assume that the thermal stress and yield stress are independent variables, so that  $p(\sigma_{th},\sigma_{y})=p_L(\sigma_{th})p_H(\sigma_{y})$, where $p_L(\sigma_{th})$ and $p_H(\sigma_{th})$ are the probability distribution functions of the thermal and yield stress fitted by the expressions given above. Then the probability that a shear transformation is activated  due to thermal dilation effects can be calculated as
\begin{eqnarray}
  p(\sigma_{th}>\sigma_{y}) &=& \int_{0}^{\infty} [\int_{0}^{\sigma_{th}}p(\sigma_{y})\mathrm{d} \sigma_{y}]p(\sigma_{th})\mathrm{d} \sigma_{th} \\
  &=& 1-\frac{\gamma}{\eta}\int_{0}^{\infty}\frac{e^{\lambda(1-\cosh(\frac{\sigma_{th}}{\beta}))}}{(1+\frac{\sigma_{th}}{\eta})^{\gamma+1}}d \sigma_{th}
\end{eqnarray}
Using the parameters extracted from simulation data, we can then estimate  this probability numerically.  The results are  shown in table \ref{tab:2}, and discussed in the following section.
\begin{table}
  \centering
  \begin{tabular}{|c|c|c|c|c|c|c|}
  \hline
  Sample & $\beta(\text{GPa})$ & $\lambda$ & $\gamma$ & $\eta(\text{GPa})$ & $p(\sigma_{th} > \sigma_{y})$ \\
  \hline
  $10^{9}$ K/s & 1.56 & 0.05 & 6.00 & 0.411 & 0.00018\\
  \hline
  $10^{13}$ K/s & 1.61 & 0.10 & 8.15 & 0.72 &  0.00046\\
  \hline
\end{tabular}
  \caption{The parameters of the Lomax distribution $p_L$ for the thermal stresses and of the hyperbolic distribution $p_H$ for the yield stresses, for  samples with different thermal histories. And $\beta$ and $\lambda$ are fitted in Fig.\ref{fig:3}, $\eta$ and $\gamma$ are fitted in Fig.\ref{fig:2}, where $\Delta T$ = 100 K}
  \label{tab:2}
\end{table}
\section{discussion and conclusion}
\begin{figure}[!htpb]
  \begin{center}
	\includegraphics[width=1.0\textwidth]{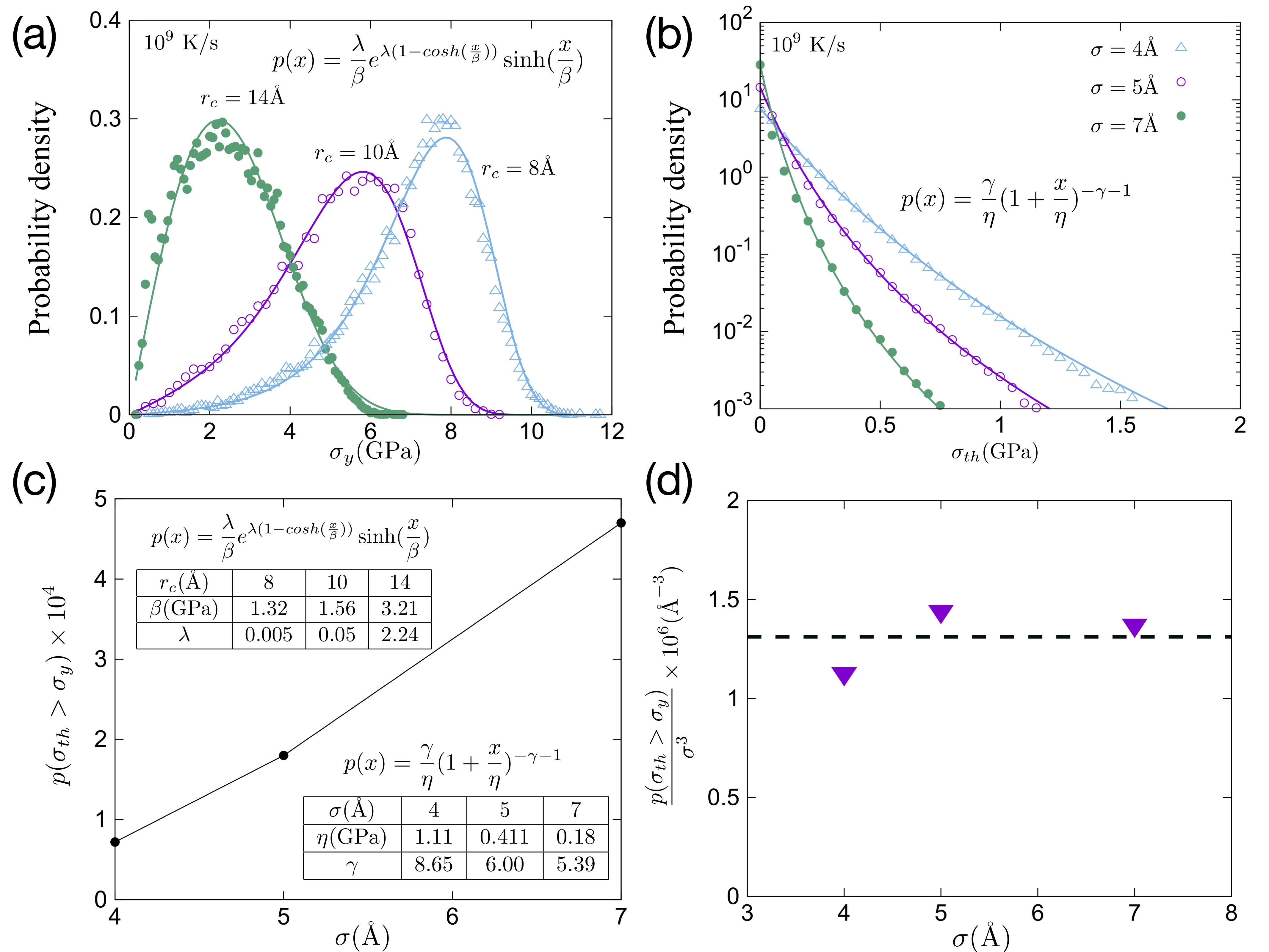}
  \end{center}
  \caption{(a) distribution of local yield stress for $r_c= 8\text{ \AA}$(hollow triangle),$r_c= 10\text{ \AA}$(hollow circle), $r_c= 14 \text{ \AA}$(solid circle), respectively, the solid line is hyperbolic probability distribution. (b) distribution of thermal stress for $\sigma=4 \text{ \AA}$, $\sigma= 5\text{ \AA}$,$\sigma=7 \text{ \AA}$, the coarse grain sizes for local yield stress and thermal stress follow the relation $r_{c} = 2 \sigma$ , and the solid line is fitted by Lomax distribution, $\Delta T=100 \text{ K}$ (c) the plastic active probability for different coarse grain size and  in the panel parameters used in hyperbolic probability distribution and Lomax probability distribution which is used to calculate the plastic active probability (d) the ratio $\frac{p(\sigma_{th}>\sigma_{y})}{\sigma^{3}}$ for different coarse grain size, the dash line is guided for the eye, and the value is the mean of three coarse grain size which equals $1.31 \times 10^{-6} \text{ \AA}^3$. The sample quench rate is $10^{9}\text{ K/s}$.}
\label{fig:app2} 
 \end{figure}
   The probability obtained in table \ref{tab:2} is small. On the scale of a simulation box that contains a few thousands  atoms (or equivalently, a few hundred possible shear transformations), the chance of actually observing a plastic event triggered by thermal stresses will be negligible. On the other hand, this probability will be significant on the macroscopic scale of an experiment. More precisely, if we consider a sample divided in  $n$ potential shear transformation sites, the probability of observing at least one shear transformation in this sample during one temperature cycle is 
   $P_{\text{act}}=1-(1-p(\sigma_{th}>\sigma_{y}))^{n}$, which we will refer to as the activation probability. This probability becomes larger than $90\%$ for  $n \approx 5000$. This corresponds to a system of lateral size that can be roughly estimated to  10 nm, assuming that the system is made of non overlapping  zones that can undergo independent shear transformations. 
   
   An important question concerns the robustness of our results with respect to the particular choice of coarse graining size. While our initial choice was motivated by the typical scale of shear transformations in metallic glasses, it remains somewhat arbitrary. Moreover, it is seen 
   in figure \ref{fig:1} (b), that the variance of the local thermal expansion  depends on coarse grain size $\sigma$ as a power law, so actually there is no characteristic coarse grain size for local thermal expansion. We therefore determined  $P_{\text{act}}$ for different coarse grain size, with the idea that this activation  probability is a physical value in a given sample, and  should not be sensitive to  the arbitrary  coarse graining size $\sigma$. Since  $p(\sigma_{th}>\sigma_{y}) \ll 1$ we can in general simplify the expression of the activation probability,  $P_{\text{act}} \sim n p(\sigma_{th}>\sigma_{y})$, where $n$ is the number of non overlapping coarse grain zones . 
 For a given system, we have $n \sim \frac{1}{\sigma^{3}}$. Therefore, the assumption that $P_{\text{act}}$ is a physical quantity implies that $p(\sigma_{th}>\sigma_{y})$ should increase with coarse grain size, while $P_{\text{act}} \sim \frac{p(\sigma_{th}>\sigma_{y})}{\sigma^{3}}$ should remain constant.  In Fig.\ref{fig:app2}, we show the results obtained for three different coarse graining sizes to test this hypothesis. As  the coarse graining size increases, the  distributions of the local yield stress and of the thermal internal stress are modified, with typically higher probabilities to obtain low values of the stress. The balance between the two evolutions leads to a nontrivial change in  $p(\sigma_{th}>\sigma_{y})$, which  increases with coarse grain size. In Fig.\ref{fig:app2}(d), we see that this increase leads indeed to a value of $\frac{p(\sigma_{th}>\sigma_{y})}{\sigma^{3}}$ that  just fluctuates around its average value, therefore establishing the physical character of   $P_{\text{act}}$.
 
 Two factors will influence the efficiency of the cryogenic rejuvenation effect. One is the number of cycles, which will progressively increase the density of events until a significant rejuvenation is achieved, bringing the system into a higher energy state. 
Indeed, in  the experiments \cite{ketov2015rejuvenation}, it was observed that the sample should be trained for several thermal cycles to obtain a notable rejuvenation effect.  Eventually this rejuvenation will be counterbalanced by the aging relaxation from the higher energy states, until a stationary situation is reached. 
 
The second important factor is of course the amplitude $\Delta T$ of temperature cycles.
Following  equation \ref{eqn:7}, the internal thermal stress increases in proportion  of the  amplitude of the temperature cycle and the local yield stress is insensitive with temperature at low temperature, thus $p(\sigma_{th}>\sigma_{y})$ would increase with the temperature amplitude, so that one could expect to amplify the effect by increasing $\Delta T$.
However, when the temperature increases, thermal relaxation will come into the picture and eventually dominate the structural evolution.  Aging effects would then hide  the rejuvenation effect caused by thermal expansion  heterogeneity, and large  thermal cycling would induce an aging effect\cite{Wang2013effect,ketov2018cryothermal}. Depending on specificities of each sample (thermal history, time scale of thermal relaxation) different quantitative effects may be observed. In  conclusion, our calculation provides the first quantitative evidence that thermally induced internal stresses have the potential for activating mechanically the plastic activity in a glassy sample, and it provides an atomic mechanism for thermal cycling rejuvenation.


\section{acknowledgments}
This work is supported by (BSS and PGF) the NSF of China (Grant Nos.51601009,Nos.51571011),the MOST 973 Program (No.2015CB856800) and the NSAF joint program (No.U1530401).  BSS and PFG acknowledges the computational support from the Beijing Computational Science Research Center (CSRC). J-L. Barrat is supported by Institut Universitaire de France 
\section*{Appendix}
\renewcommand{\thefigure}{A\arabic{figure}}
\renewcommand{\theequation}{A\arabic{equation}}
\setcounter{figure}{0}
\setcounter{equation}{0}
Following the coarse grain method in ref \cite{PhysRevE.80.026112}, we calculated local elastic modulus with $\sigma=5\text{ \AA}$, and using the same spatial position as LTEC, the distribution and correlation are shown in Fig.\ref{fig:app1}. Clearly, there is no correlation between local thermal expansion and local elastic moduli.
\begin{figure}[!htpb]
  \begin{center}
	\includegraphics[width=1.0\textwidth]{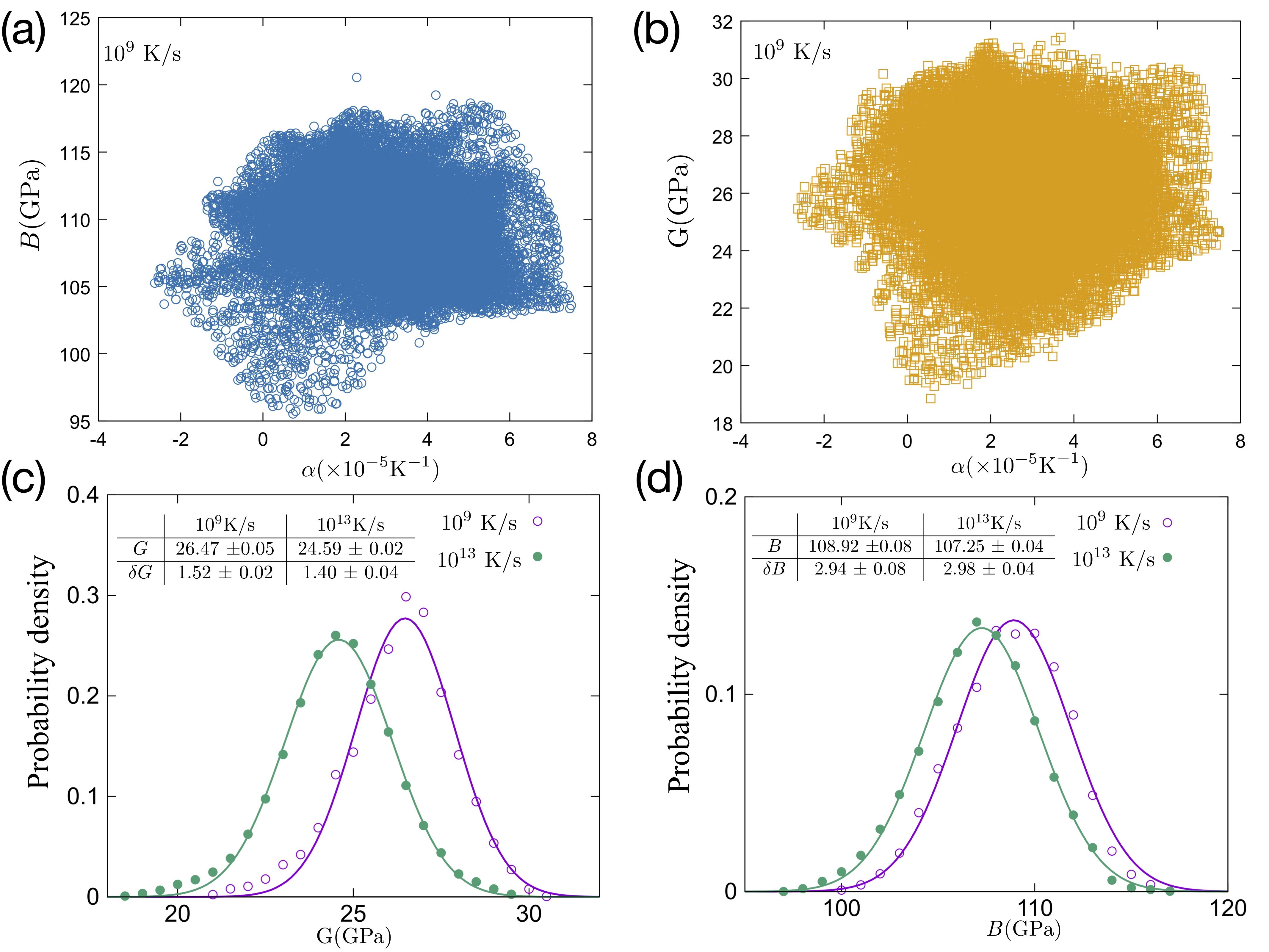}
  \end{center}
  \caption{(a),(b) scatter plot of the  LTEC versus local modulus : bulk(B),shear(G) respectively. (c),(d) distribution of local modulus (bulk and shear) for different thermal histories of the  sample :$10^{9}$ K/s,$10^{13}$ K/s respectively, the parameters of the fit to a Gaussian  are shown in the inset of (c),(d).}
\label{fig:app1} 
 \end{figure}
 Next, we show that, as a first order approximation for the heterogeneity, our uniform elastic constant assumption in the article is reasonable.
 
   We take equation \ref{eqn:7} in the article as an illustration, and introduce the possibility of heterogeneous elasticity in this equation. A similar reasoning can be extended to the complete calculation. In the presence of heterogeneity, we can rewrite equation \ref{eqn:7} as:  
   \begin{equation}
\partial_{j} (K(r)+\frac{4}{3}G(r))\partial_{j} \Phi+K(r)\Delta T(\alpha(r)-\bar{\alpha})=0
\label{eqn:A1}
\end{equation}
 we use the variance and mean of the local moduli to represent the heterogeneity, as $K(r)=\bar{K}+\delta K; G(r)=\bar{G}+\delta G; \alpha(r)-\bar{\alpha}=\delta \alpha$.
 For a uniform medium, the solution is obtained by inverting the linear operator $\partial_{j} (K(r)+\frac{4}{3}G(r))\partial_{j} $ (which is conveniently done in Fourier space) and applying its inverse $L_0$ to $K(r)\Delta T(\alpha(r)-\bar{\alpha})$. The presence of heterogeneities makes the corresponding operator, $L_0+\delta L$ non diagonal in Fourier space. However, we expect that generically $L_0^{-1}\delta L $ is of the same order of magnitude as $\delta K/K$ or $\delta G/G$.
 As shown in the panel of Fig.\ref{fig:app1}, $\delta K/K$ and $\delta G/G$ are small compared to one.
 
 Hence,  the approximation used in the manuscript can be considered as a first order approximation that ignores terms of order $\delta \alpha \times\delta G/G $ or  $\delta \alpha \times\delta K/K $. 
\bibliographystyle{unsrt}
\bibliography{bibalpha}
\end{document}